# Synthesis and properties of La-doped $CaFe_2As_2$ single crystals with $T_c$ = 42.7 K

Zhaoshun Gao[*], Yanpeng Qi[*], Lei Wang, Dongliang Wang, Xianping Zhang, Chao Yao, Chunlei Wang, Yanwei Ma[1]

Key Laboratory of Applied Superconductivity, Institute of Electrical Engineering, Chinese Academy of Sciences, Beijing 100190, China

**Abstract:**

Large single crystals of La-doped $CaFe_2As_2$ were successfully synthesized by the FeAs self-flux method. The x-ray diffraction patterns suggest high crystalline quality and c-axis orientation. By substitution of trivalent $La^{3+}$ ions for divalent $Ca^{2+}$, the resistivity anomaly in the parent compound $CaFe_2As_2$ is completely suppressed and a superconducting transition reaches the value of 42.7 K, which is higher than that of about 30K reported in Saha S. R. et al., arXiv:1105.4798v1. The upper critical field has been determined with the magnetic field along *ab*-plane and *c*-axis, yielding an anisotropy of about 3.3.

---

[*] Joint first authors. Both authors contributed equally to this work.

[1] Author to whom correspondence should be addressed; E-mail: ywma@mail.iee.ac.cn

The discovery of superconductivity in LaFeAsO$_{1-x}$F$_x$ with critical temperature ($T_c$) 26 K has triggered enormous attention as a new family of high temperature superconductors [2]. Since the original discovery, several families of iron-based layered superconductors have been discovered, commonly denoted as 1111 phase (ReFeAsO with Re = rare earth) [3-7], 122 phase (AeFe$_2$As$_2$ with Ae = alkaline earth) [8-11], 111 phase (AFeAs with A = alkali metal) [12, 13], 11 phase (FeTe or FeSe) [14], 42622 phase (Sr$_4$Sc$_2$O$_6$Fe$_2$P$_2$ et al.)[15, 16], and FeSe122 phase (AFe$_2$Se$_2$) [17-21]. The present research indicates that the layered structure with the conducting Fe-As (Fe-Se) layers should be responsible for high temperature superconductivity. Furthermore, superconductivity emerges when the antiferromagnetical ordered phase is suppressed in the parent compounds, which looks very similar to the case of cuprates.

The pristine CaFe$_2$As$_2$ has the tetragonal ThCrSi-type crystal structure with a space group I4/mmm, similar to other member of the AeFe$_2$As$_2$ family. It is a good springboard to investigating Fe-pnictide superconductivity [22]. At ambient pressure CaFe$_2$As$_2$ undergoes a transition from a non-magnetically ordered tetragonal to an antiferromagnetic orthorhombic phase [23, 24]. When a modest external pressure is applied to CaFe$_2$As$_2$, the orthorhombic phase transforms to a collapsed tetragonal structure [25]. To stabilize the collapsed phase at ambient pressures, Saha S. R. et al. [1] have employed rare earth substitution into CaFe$_2$As$_2$. It is striking that structural collapse and 45 K superconductivity were induced in rare earth doped CaFe$_2$As$_2$, which should add more ingredients to the underlying physics of the iron-based superconductors. Here we report our new results on the La-doped CaFe$_2$As$_2$, with a resistivity onset $T_c$ of 42.7 K, which is higher than the result of about 30 K in Ref [1].

The single crystals of CaFe$_2$As$_2$ and Ca$_{1-x}$La$_x$Fe$_2$As$_2$ were grown by the FeAs self-flux method. The FeAs precursor was synthesized by the reaction of Fe powder (Alfa Aesar, 99.99% in purity) and As chips (99.999%) at 500 ◦C for 10 h and then 700 ◦C for 20 h in a sealed Niobium tube. Appropriate amounts of the starting materials of FeAs, Ca and La were placed in an alumina crucible, and sealed in an arc-welded pure iron tube. The sample was put into tube furnace and heated to 1200

°C slowly and held there for 5 hours, and then was cooled to 1030 °C with a rate of 3 ~ 6 °C/h to grow the single crystals. The obtained single crystals show the shiny surface and are easily cleaved into plates, as shown in inset of Fig. 1.

The phase identification and crystal structure investigation were carried out using x-ray diffraction (XRD). The composition of crystals was determined by using an Energy Dispersive X-ray Spectrometer (EDXS). Standard four probe resistance and magnetic measurements were carried out using a physical property measurement system (PPMS). Rectangular specimens with dimensions of about $4 \times 2 \times 0.2$ mm$^3$ were cut from the samples and resistivity measurements were performed by the conventional four-point-probe method. The electric contacts were made using silver paste with a baking step for 15 min at 120 °C. Due to the large thermal expansion coefficient of $Ca_{1-x}Re_xFe_2As_2$, the baking step is essential for a good electric contact.

The crystal structure of $CaFe_2As_2$ and $Ca_{0.8}La_{0.2}Fe_2As_2$ samples were examined by x-ray diffraction measurement on the *ab*-plane of the crystals. Typical diffraction patterns are shown in figure 1. Only sharp 00*l* (*l* = 2n) reflections were recognized, indicating that the single crystals were perfectly oriented along c-axis. It is found that the position of (00*l*) peaks of La doped sample shifts to higher angles, suggesting a successful chemical substitution and decrease in *c*-axis lattice. The EDX analyses show that the actual compositions of the single crystals are close to the nominal ones.

Fig. 2 displays the temperature dependence of the normalized resistance for single crystals of $CaFe_2As_2$ and $Ca_{0.8}La_{0.2}Fe_2As_2$ respectively. The parent $CaFe_2As_2$ compound exhibits a resistivity anomaly at about 165 K, which associated with the magnetic/structural phase transition. The anomaly is completely suppressed and a superconducting transition reaches the value of 42.7 K in La doped sample $Ca_{0.8}La_{0.2}Fe_2As_2$. It suggests that partial substitution of La for Ca induces electron type carrier into system, and leads to the suppression of magnetic and structural instabilities and induces superconductivity in $Ca_{0.8}La_{0.2}Fe_2As_2$. The superconductivity of our sample is also confirmed by DC magnetization measurement which is shown in Fig. 3. A clear diamagnetic signal appears around 40 K in the samples, which correspond to the middle transition temperature of the resistivity data. The relatively broad magnetic transition may be ascribed to the local inhomogeneous of the stoichiometry. This kind of broad transition in doped $CaFe_2As_2$ samples was also reported [1, 26].

The temperature dependence of resistivity with various magnetic fields applied along *ab*-plane and *c*-axis were presented in Fig. 4 (a) and (b). It can be seen that the onset of $T_c$ decreases with increasing magnetic field very similar to $T_c^{zero}$ for both *H//ab* and *H//c*, but the effect of magnetic field is much larger when the field is applied along the c axis of the single crystals. It is worth to note that the transition temperature is sensitive to magnetic field under the condition of *H//c*. The temperature transition shifts very large to lower temperature even under 0.1 T. The upper critical fields ($H_{c2}$) of single crystal $Ca_{0.8}La_{0.2}Fe_2As_2$ were determined using a criterion of 90% points on the resistive transition curves. The estimated upper critical fields were plotted in Fig. 5 as a function of temperature. It is clear that the curves of $H_{c2}$ (T) are steep with the slopes of $-dH_{c2}/dT|_{Tc}$ = 6.25 T/K for *H//ab* and $-dH^c_{c2}/dT|_{Tc}$ = 1.88 T/K for *H//c*. According to the Werthamer-Helfand-Hohenberg formula [27], $H_{c2}(0) = 0.693 \times (dH_{c2}/dT) \times T_c$. Taking $T_c$= 42.7 K, we can get the values of upper critical fields close to zero temperature limit: $H^{ab}_{c2}$= 185 T and $H^c_{c2}$= 55.6 T. We should note that the $H_{c2}$ values derived by conventional single-band WHH approximation may differ from the actually measured $H_{c2}$ values. However, our results clearly indicate that the $H_{c2}$ values in our samples are really high. The anisotropy parameter of $H^{ab}_{c2}/H^c_{c2}$ is about 3.3, which is similar to that of $AFe_2Se_2$ [21, 28].

In summary, we have synthesized large $Ca_{0.8}La_{0.2}Fe_2As_2$ single crystals with the onset superconducting transition temperature 42.7 K. By substitution of La for Ca, the magnetic/structural phase transition in the parent $CaFe_2As_2$ compound is suppressed and superconductivity emerges. A high upper critical field $H_{c2}^{ab}(0)$ of 185 T was obtained. The anisotropy of the superconductor determined by the ratio of $H^{ab}_{c2}/H^c_{c2}$ is about 3.3, which is similar to that of $AFe_2Se_2$.

**Acknowledgement**

The authors thank Profs. Haihu Wen, Liye Xiao and Liangzhen Lin for their help and useful discussions. This work is partially supported by the National '973' Program (Grant No. 2011CBA00105) and National Natural Science Foundation of China (Grant No. 51002150 and 51025726).

# Captions

Figure 1   X-ray diffraction patterns of the single crystal $CaFe_2As_2$ and $Ca_{0.8}La_{0.2}Fe_2As_2$. Inset shows the photography of the single crystals (length scale 1 mm).

Figure 2   Temperature dependence of resistivity for $CaFe_2As_2$ and $Ca_{0.8}La_{0.2}Fe_2As_2$ at zero field up to 300 K.

Figure 3   Temperature dependence of dc magnetization for ZFC and FC processes at a magnetic field of H = 20 Oe.

Figure 4   (a) and (b) show temperature dependence of resistivity for $Ca_{0.8}La_{0.2}Fe_2As_2$ single crystal with the magnetic field parallel to the *ab*-plane and *c*-axis up to 9 T.

Figure 5   The upper critical fields of $Ca_{0.8}La_{0.2}Fe_2As_2$ single crystal for magnetic fields parallel to the ab-plane and c-axis respectively.

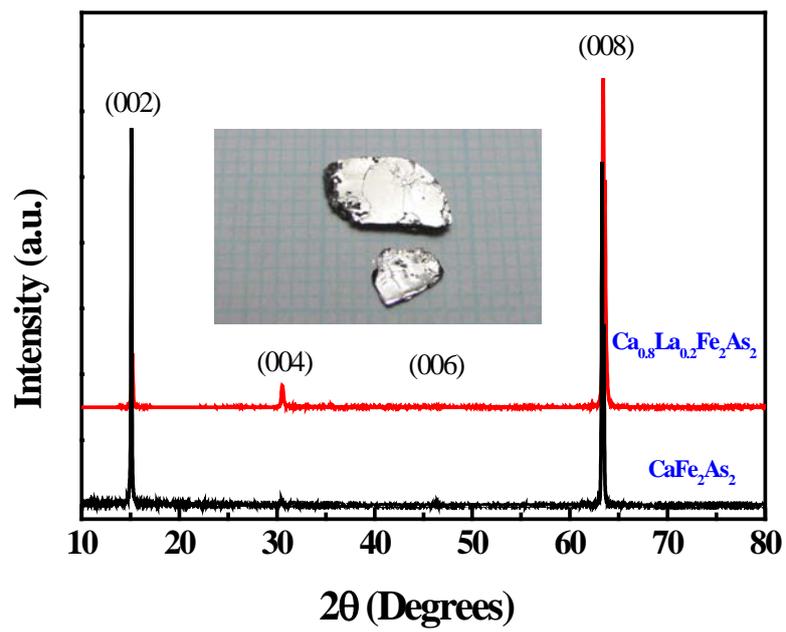

Fig.1 Gao et al.

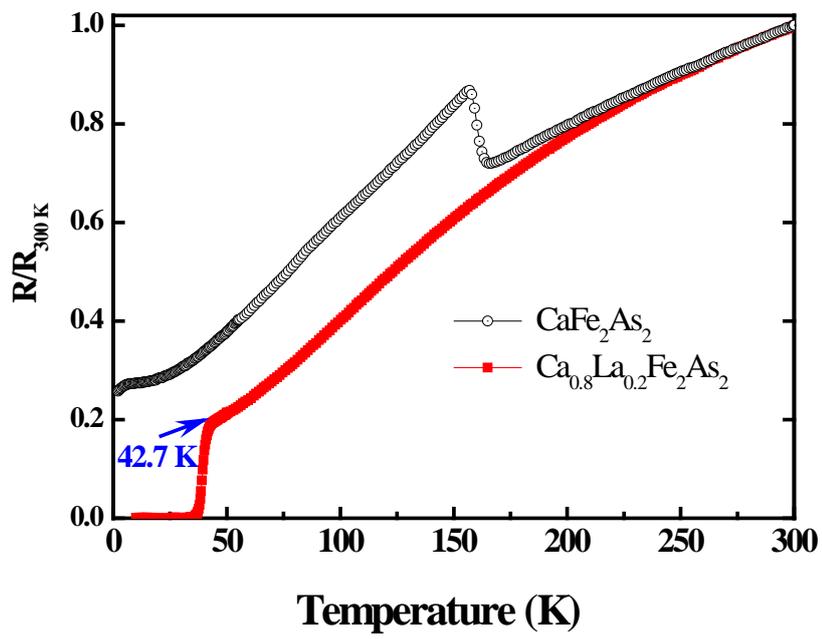

Fig.2 Gao et al.

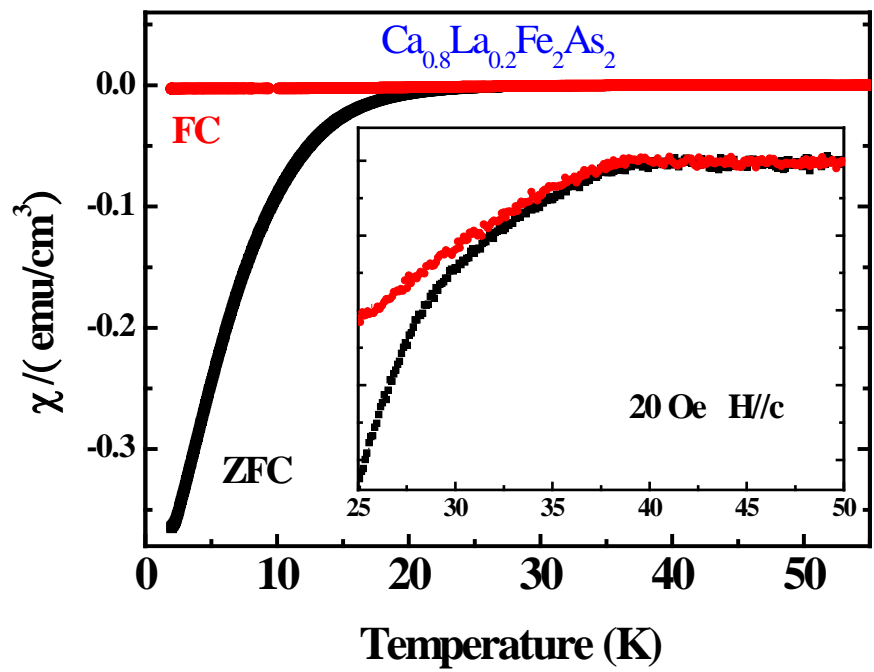

Fig.3 Gao et al.

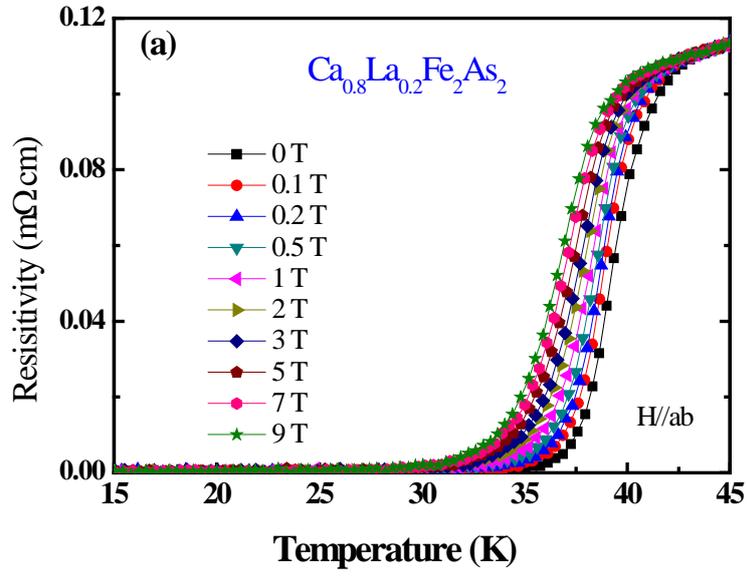

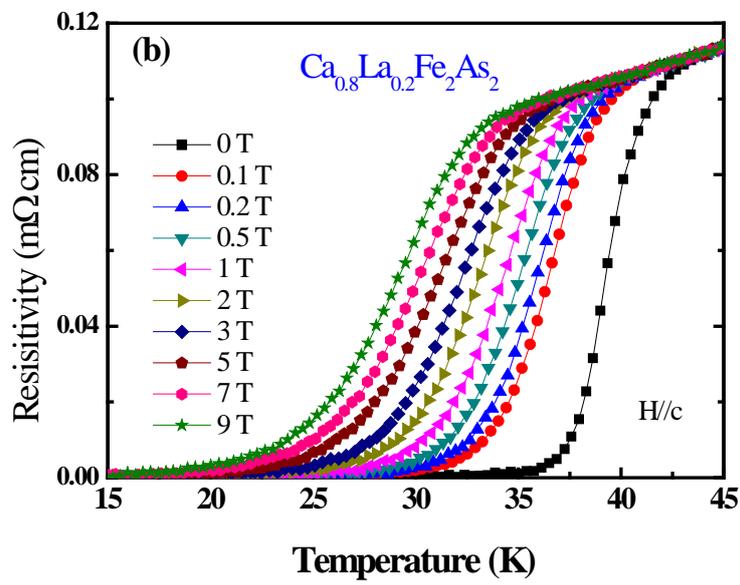

Fig.4 Gao et al.

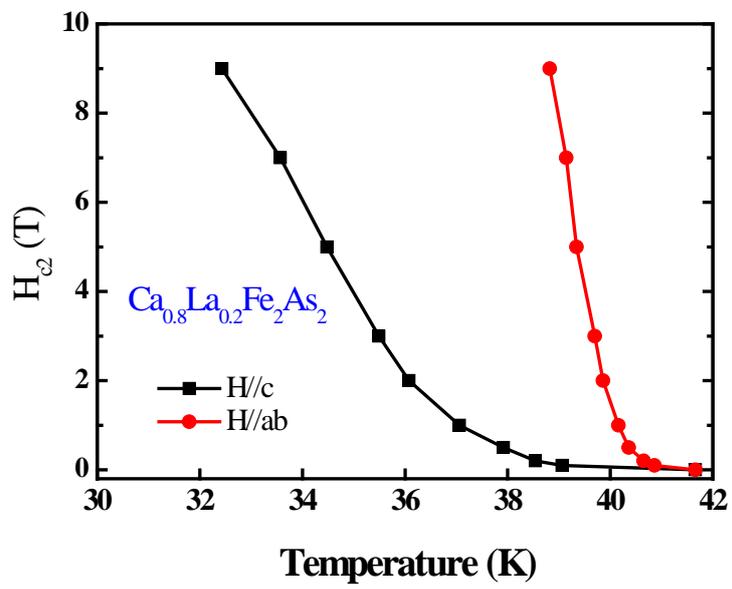

Fig.5 Gao et al.